\documentclass[preprint,preprintnumbers,amsmath,amssymb]{revtex4}
\usepackage{graphicx}
\def\alf{Alfv\'en\,}
\def\bq{\begin{equation}}
\def\eq{\end{equation}}
\let\grad=\nabla

\def\v{{{\bf{v}}}}

\def\vi{{{\bf{v}}}_i}
\def\vj{{{\bf{v}}}_j}
\def\ve{{{\bf{v}}}_e}
\def\vi{{{\bf{v}}}_i}

\def\vB{{{\bf{v}}}_B}
\def\va{v_A}
\def\vn{{{\bf{v}}}_n}
\def\B{{\bf{B}}}

\def\hB{\hat{{\bf{b}}}}

\def\J{{{\bf{J}}}}
\def\Jpa{\J_{\parallel}}
\def\Jpe{\J_{\perp}}  
\def\E{{{\bf{E}}}}
\def\Ep{{\bf{E'}}}
\def\Epa{\Ep_{\parallel}}
\def\Epe{\Ep_{\perp}}  
\def\k{{{\bf{k}}}}

\def\x{{{\bf{x}}}}

\newcommand\cross{{\bf{\times}}}
\def\curl{{\grad \cross}}
\newcommand{\delt} [1] {\frac{\partial #1}{\partial t}}

\begin{document}
 
\title{\bf The magnetohydrodynamic description of Earth{}\'s ionosphere}
\author{\bf B. P. Pandey}
\affiliation{Department of Physics and Astronomy, Macquarie University, Sydney, NSW 2109,
Australia.}
\email{birendra.pandey@mq.edu.au}

\begin{abstract}
The wave propagation in the partially ionized ionosphere plays an important role in the magnetosphere—-ionosphere coupling. For example, the ionosphere may supports very low-frequency \alf wave which can be caused by a balance between the bulk fluid inertia (mostly due to neutrals in the lower and middle E--region) and the deformation of the magnetic field. The plasma-—neutral collisional momentum exchange facilitates the transfer of the magnetic stress (felt directly by the ions) to the neutrals. Therefore, in the low-frequency (with respect to the neutral-—ion collision frequency) limit, waves through the ionosphere can propagate with very little damping. In the vanishing plasma inertia limit, waves can be excited due to the {\it loading} of neutral inertia on the field lines and thus may have very long wavelength and can easily couple to the magnetosphere. The frequency of these waves are below few Hz.

In the present work dynamics of the neutral particles have been included in the  single—-fluid, ideal magnetohydrodynamic like framework where owing to the frequent collisions between the neutral and plasma particles, the magnetic field is no longer frozen in the partially ionized fluid but slips through it. The relative importance of this non—-ideal behaviour of the fluid is dependent on how often the plasma particles suffer collision with the neutral particles. In the limiting case, when the non—-ideal effects can be ignored, the equations for both the partially and fully ionized plasmas are identical except inertia of the partially ionized fluid is carried by both the ions and the neutrals.
\end{abstract}

\maketitle

\pagebreak

\section{Introduction}
The Earth{}\'s lower ionosphere ($\lesssim 400\,\mbox{km}$) poses considerable challenge to the development of an unified framework for the magnetosphere-—ionosphere (MI) coupling. The origin of this difficulty lies in the varying fractional ionization of the matter in the ionosphere and the magnetosphere. The ionosphere is in a partially ionized plasma state and is highly collisional whereas the magnetospheric ($> 400\,\mbox{km}$) plasma {\it to a large extent} can be treated as a collisonless magnetohydrodynamic (MHD) fluid [1]. This results in different approaches taken to study the ionosphere and magnetosphere. Whereas collisonless magnetospheric plasma is generally described in the ideal magnetohydrodynamics (MHD) framework, i.e. in terms of bulk plasma flow ($\v$) and magnetic field ($\B$)-—so called $\left(\v, \B\right)$ paradigm, the ionospheric dynamics is described in terms of current ($\J$) and electric field ($\E$), so called $\left(\E, \J\right)$ paradigm [2,3]. However, the very first serious difficulty of the $\left(\E, \J\right)$ paradigm is conceptual. Since the electric field in the ionosphere is assumed to have its origin in the {\it external} events (e.g. solar wind [4] and references therein), it cannot penetrate the quasi--neutral ionospheric plasma except in a very thin boundary layer of the order of the Debye Length. Therefore, the limited penetration of such an electric field in the plasma will not be able to set the convective motion and drive currents [5]. Another difficulty with the $\left(\E, \J\right)$ paradigm is that the MI coupling is achieved by determining the current density from the stress balance and the electric potential is assumed constant along the field line. However, such a description of the ionosphere and the resultant coupling to the magnetosphere not only overlooks the dynamics of the neutral particles but also presupposes stationary conditions [2\,,3\,,5].

The observations of the mid latitude E—-region of the ionosphere suggest the presence of the very low frequency ($\lesssim 10^{-4}\,\mbox{s}^{-1}$ long wavelength ($10^3\,\mbox{km}$ slow planetary waves with a period varying between several days to less than a day [6] and fast waves with a period varying between several tenth of a minute to about a minute [7]. In the F-—region of the ionosphere the fast planetary electromagnetic waves with the phase velocities from tens to several hundred km/s having similar wavelength and periods like in the E—-region are also observed [8]. The phase velocities of the waves vary by an order of magnitude from the day to night in the E-—region. Owing to their large phase velocities and their strong day—-night phase variations in the E—-region, it is impossible to identify these waves as the ordinary magnetohydrodynamic (MHD) waves. Although it has long been recognised that the low frequency, long wavelength signals which propagate through the ionosphere plays an important role in the MI coupling [6], the contribution of the neutral atmosphere to the dynamics has only recently been recognised [9].  The inclusion of the neutral dynamics in a self-—consistent framework tantamount to including the very low frequency dynamical response of the medium. Like ref.~[9] we aim to develop a self-—consistent single fluid  
description of the partially ionized ionosphere which as we shall see reduces to the usual MHD description in the limiting case. In such a framework, the magnetic, collisional and pressure forces interacts with each other within extremely collisional environment at the lower end of the ionosphere and within the perfectly conducting environment at the higher end of the atmosphere.  The advantage of such an approach is that the effect of the neutral dynamics on the signal propagation through ionosphere can be properly accounted for. Note that the fluid description has its own range of validity set mainly by the relevant length and time scales.     

The partially ionized ionospheric plasma is a mixture of electrons, ions and neutral particles. The three fluid equations provides a detailed description of such a plasma [10, 11, 12].   Whereas multi-fluid framework is well suited to describe both the high and low--frequency fluctuations in various limits, it is still mathematically cumbersome and realistic treatment of the system is formidable. Thus it is desirable to further reduce the three fluid equations to a single fluid, MHD--like description which not only provides the economy of description but also is much simpler to handle. However, like a single fluid MHD description of a fully ionized two component plasma, the single fluid MHD—-like description of a partially ionized three component plasma will be suitable for the investigation of only the very low-—frequency, long wavelength response of the medium. Thus the single fluid description (Ref.~[13]; hereafter PW08), will be valid only when the plasma and the neutral particles are glued together by the frequent collision and thus move together  as a single fluid, i.e. when
\bq
    \left(\frac{\omega}{\nu_{ni}}\right) \lesssim \left(\frac{n_e}{n_n}\right)^{-1/2} \equiv X_e^{-1/2} \,.
\label{eq:omega1}
\eq
Here $X_e = n_e/n_n$ is the fractional ionization, $\omega$ is the signal frequency, and $\nu_{ni}$ is the neutral--ion collision frequency which is related to the ion--neutral collision frequency $\nu_{in}$ by the following relation $\rho_n\,\nu_{ni} = \rho_i\,\nu_{in}$. Note that the above equation has been derived from Eq.~(20) of PW08 in the weakly ionized limit assuming that the bulk mass density ($\rho_j = m_j\,n_j$ where $m_j$ and $n_j$ corresponds to the mass and number density of the $j^{\mbox{th}}$ particle) of the fluid is mainly due to the neutral gas, i.e. $\rho \approx \rho_n$, or, $D = \rho /\rho_n \approx 1$. Further, electrons are assumed strongly magnetized and we have assumed $m_i \approx m_n$ (although the ion and the neutral masses in the ionosphere are different ) and imposed plasma quasi—neutrality condition $n_i \approx n_e$. These are plausible assumptions and thus Eq.~(1) provides a legitimate bound on the single fluid MHD—-like framework for the weakly ionized plasma dynamics germane to the lower and middle ionosphere. In the upper E and F regions where $X_e \gtrsim \mathcal{O}(1)$, one should use Eq. ~(20) of PW08 without above assumptions.
\begin{figure}
\includegraphics[scale=0.45]{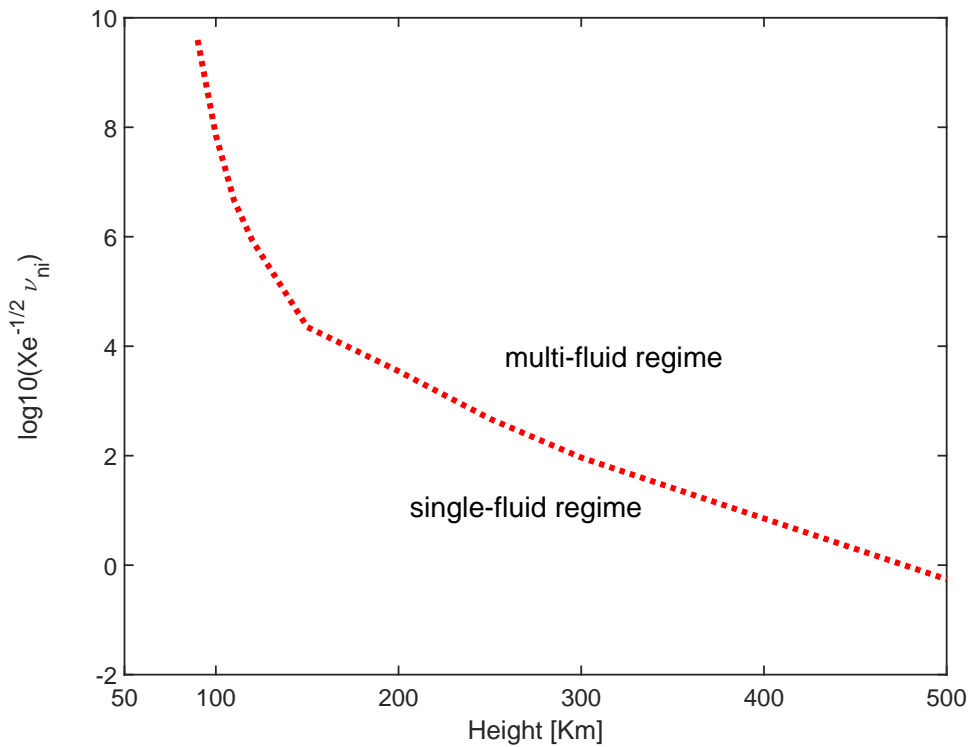}
\caption{The regions where single-fluid formulation of the multicomponent
plasma is valid for the parameters pertaining to the ionosphere.
}
\label{fig:SFV}
\end{figure}

In Fig.~({\ref{fig:SFV})  $X_e^{-1/2}\,\nu_{ni}$ is plotted against the height. It is clear from the figure that at mid altitude ($\lesssim 150-160\,\mbox{km}$)  the single fluid description is valid for the signal speed of the order of $\omega \lesssim 10^4\,\mbox{HZ}$. In the upper ionosphere $\gtrsim 300\,\mbox{km}$ the dynamics of low frequency ($\lesssim 10\,\mbox{Hz}$) signals can be described in the single fluid framework. For example, the generation and propagation of the ultra—low ($10^{-3}\,\mbox{Hz}$) frequency planetary waves ($\sim 10^3\,\mbox{km}$) can be investigated in this framework. To summarize, the single fluid MHD-—like description is valid in both the ionosphere and magnetosphere if the signal frequency under investigation is very low or, the wave length is very large-—a case investigated in ref.~[9].  

How the present single fluid formulation of the partially ionized plasma does differs from the classical single fluid MHD description of the fully ionized plasma?  They differ on several counts. (a) The fluid inertia in a partially ionized medium is due to the ions as well as the neutrals. Thus at the lower and mid altitude, where plasma is weakly ionized, the inertia of the fluid is due mainly to the neutrals whereas at the higher altitude mostly positive ions provides the fluid inertia. (b) Therefore, the \alf wave propagation in the medium with varying fractional ionization changes its character with altitude: in the lower ionosphere, it is the neutral inertia that balances the magnetic field deformation while at higher altitude, the ion inertia balances the magnetic field fluctuations. (c) The wave propagation is affected by the frequent collisions between the plasma and the neutral particles. These collisions allow the plasma and neutral particle to move together as a single fluid if the wave frequency satisfies Eq.~(\ref{eq:omega1}). (d) However, these collision will cause the finite drift/slippage of the magnetic field through the partially ionized medium. (e) When the magnetic field drift is ignored, the characteristics of the wave propagation in the partially ionized fluid is similar to the fully ionized ideal--MHD except now the propagation properties depend on the fractional ionization as well as on the compressibility of the medium.
  
In this paper, we first give a very brief outline of the single fluid derivation in section II. The detailed discussion about the validity of the single fluid formulation is given elsewhere [PW08]. The variation of non—-ideal MHD effects such as Ohm, Hall and ambipolar diffusion in the ionosphere and their relative importance with respect to the advection term (in terms of various Reynold numbers) is also discussed. We use these equations in section III to investigate the low frequency wave propagation and their attenuation in the ionosphere. In section IV  the discussion and summary of  results is give.

\section{ Basic Model}
The partially ionized ionospheric plasma can be described in a multi—fluid framework, where the electrons, ions and neutral particles are described by the three fluid equations [9\,,10\,,11\,,12]. The continuity equation is
\begin{equation}
\frac{\partial \rho_j}{\partial t} + \grad\cdot\left(\rho_j\,\vj\right) = 0\,,
\label{eq:contj}
\end{equation}
where $\vj$ is the velocity of various components, $j=i,\,e,\,n$.  We shall assume that the ions are singly charged. The momentum equations for the electrons, ions and neutrals are
\begin{equation}
\frac{d\ve}{dt}= - \frac{\nabla P_e}{\rho_e} - \frac{e}{m_e}\left(\E + \frac{\ve}{c}\cross \B\right)
-\!\! \sum_{j=i,n}\!\!\nu_{ej}\left(\ve - \vj \right)
\label{eeq}
\end{equation}
\begin{equation}
\frac{d\vi}{dt}=   -\frac{\nabla\,P_i}{\rho_i}  + \frac{e}{m_i}\,\left(\E + \frac{\vi}{c}\cross \B\right)
-\!\! \sum_{j=e,n}\!\!\nu_{ij}\left(\vi - \vj \right)
\label{ieq}
\end{equation}
\begin{equation}
\frac{d\vn}{dt}= - \frac{\nabla\,P_n}{\rho_n} +
\sum_{j=e,i} \!\!\nu_{nj}\left(\vj - \vn \right)\,.
\label{neq}
\end{equation}
The electron and ion momentum equations (\ref{eeq})-(\ref{ieq}) contain on the right hand side pressure gradient, Lorentz and collisional momentum exchange terms where $P_j$ is the pressure, $\E$ and $\B$ are the electric and magnetic field, $c$ is the speed of light, and $\nu_{ij}$ is the collision frequency for species $i$ with species $j$.  The electron-ion collision frequency $\nu_{ei}$ can be expressed in terms of fractional ionization $X_e = n_e/n_n$  and plasma temperature $T_e = T_i = T$ as [PW08]
\begin{equation}
\nu_{ei} = 51\,X_e\,n_n\,T^{-1.5}\,\mbox{s}^{-1}\,,
\end{equation}
where $\mbox{T}$ and $n_n$ are in $\mbox{K}$ and $\mbox{cm}^{-3}$ respectively. 
The ion--neutral  $\nu_{i n}$ and electron—-neutral collision frequencies are [14]
\begin{eqnarray}
\nu_{i\,n} = 2.6 \times 10^{-9}\,A^{- 1 / 2}\,n_n\,,\nonumber\\
\nu_{e\,n} = 5.4 \times 10^{-10}\,T^{1 / 2}\,n_n\,,
\end{eqnarray}
where $A \approx 28$ is the mean neutral mass in the atomic mass unit.

Defining bulk fluid density $\rho \approx \rho_i + \rho_n$, and velocity $ \v = (\rho_i\,\vi + \rho_n\,\vn)/\rho$, Eqs.~(\ref{eq:contj})-—(\ref{neq}) can be reduced to a single fluid description. These equations are 
\begin{equation}
\frac{\partial \rho}{\partial t} + \grad\cdot\left(\rho\,\v\right) = 0\,.
\label{eq:continuity}
\end{equation}
\begin{equation}
\rho\,\frac{d\v}{dt} = - \nabla\,P + \frac{\J\cross\B}{c} \,,
\label{eq:meq1}
\end{equation}
where $P = P_e + P_i + P_n$ is the total pressure, and $\J = n_e\,e\,\left(\vi - \ve\right)$ is the current density. Note that the fractional ionisation varies between $10^{-9}-–10^{-4}$ at mid and high altitude, and thus the pressure is largely due to the neutral component. 

The induction equation is derived from the electron momentum equation.  
We write the induction equation in the ideal MHD form so that it facilitates the visualisation of the magnetic field as {\it a real physical entity} that is drifting through the fluid with a given velocity. Here by a real physical entity we mean that {\it if the fluid particles are on the same line of force at any time, then they will always be on that line of force} [15]. In this {\it frozen--in} frame, with the field and fluid drifting with $\vB$, we have $c\,\Ep + \vB\cross\B = 0$. Here 
\bq
\Ep = \E + \frac{\vn\cross\B}{c}\,,
\eq
is the electric field in the neutral frame and $\E$ is the electric field in the {\it fixed} frame of reference, commonly the frame of the rotating Earth. The induction equation can be written as [PW08]
\bq
\delt \B = \curl\left[
\left(\v + \vB\right)\cross\B - \frac {4\,\pi\,\eta_O}{c}\,\Jpa\right]\,,
\label{eq:ind1}
\eq
where 
\bq	
\vB = \eta_P\,\frac{\left(\grad\cross\B\right)_{\perp}\cross\hB}{B} -– 
\eta_H\,\frac{\left(\grad\cross\B\right)_{\perp}}{B}\,. 
\label{eq:md0}
\eq
Here $\eta_P$ and $\eta_H$ are the Pedersen and Hall diffusivities and 
$\hB = \B/B$. The Pedersen diffusivity $\eta_P$ is the sum of the Ohm ($\eta_O$) and the ambipolar ($\eta_A$) diffusion, i.e.
\bq
\eta_P = \eta_O + \eta_A\,.
\label{eq:ets}
\eq
In terms of the ambient plasma parameters various diffusivities are
\bq \eta_O =
\frac{c^2}{4\,\pi\sigma}\,,\quad \eta_{A} =
\frac{D\,v_A^2}{\nu_{ni}}
\,,\quad \eta_H = \frac{v_A^2}{\omega_{H}}\,,
\label{eq:diffu}
 \eq
where 
\bq
\sigma = \frac{e^2\,n_e}{m_e\,\left(\nu_{en} + \nu_{ei}\right)}\,,
\eq
$D = \rho_n / \rho$, $v_A = B / \sqrt{4\,\pi\,\rho}$ is the \alf speed in the bulk fluid,  
\bq
\omega_{H} = \frac{\rho_i}{\rho}\,\omega_{ci} 
\eq
is the Hall frequency [PW08] and $\omega_{ci} = e\,B/m_i\,c$ is the ion—-cyclotron frequency.  It is pertinent to note here that in the ionosphere at heights between $80—-400\,\mbox{km}$, since the fractional ionization is very low, $\rho_i/\rho \sim 10^{-9}-—10^{-4}$ the Hall frequency is $\sim 10^{-6}—-10^{-1}/\mbox{s}$. 

From the induction equation (\ref{eq:ind1}) we see that when the Ohm diffusion is negligible, the magnetic flux is frozen in a frame moving with $\v + \vB$. Thus both ambipolar and Hall diffusion only redistributes the flux in the medium. Only Ohm diffusion is capable of destroying the flux at a rate proportional to its local value in the ionosphere. The relative importance of various diffusion is closely related to the question of how well the ionized matter is coupled to the magnetic field quantified by plasma Hall parameter
\bq
\beta_j = \frac{\omega_{cj}}{\nu_j}\,,
\eq
a ratio of the $j^{\mbox{th}}$ particles cyclotron $\omega_{cj} = e\,B / m_j\,c$ and collision frequencies. 

The ratio of the ambipolar and Hall diffusion is determined by ion-Hall parameter $\beta_i$ and the ratio of Ohm and Hall diffusion is determined by the electron Hall parameter $\beta_e$, i.e.
\bq
\eta_A = D\,\beta_i\,\eta_H\,,\mbox{and}\,,\eta_O = \beta_e^{-1}\,\eta_H\,.
\label{eq:difR}
\eq
\begin{table*}
 \centering
 \begin{minipage}{140mm}
  \caption{\label{tab:table1}The ion and electron plasma beta and \alf speed in the neutral medium is shown against the height for the ionospheric parameters.}
  \begin{tabular}{@{}llll@{}}
  \hline
H[km] & $\beta_i$ & $\beta_e$ & $v_A(\mbox{km}/\mbox{s})$\\
\hline
 $100$ & $10^{-2}$ & $10^{3}$ &  $6.3\times10^{-2}$   \\
 $200$  & $3\times 10^{2}$ & $10^{4}$  & $3.6$   \\
 $400$ & $10^{4}$ & $10^{6}$ & $47$  \\
\hline
\end{tabular}
\end{minipage}
\end{table*}
The characteristic scale lengths associated with the various diffusivities are
\bq
L_H = \frac{\va}{\omega_H}\,,\quad
L_A = D\,\beta_i\,L_H \equiv \frac{\va}{\nu_{ni}}\,,\quad
L_O = \beta_e^{-1}\,L_H\,.
\label{eq:LH}
\eq
Since only the electrons are magnetized in the ionosphere $\beta_e > 1$, we can say that the scale $L_O$ over which the Ohm diffusion operates is  smallest among the three scales. As is clear from the table, in the lower ionosphere $\beta_i < 1$ and the ions follow the neutrals but the electrons follow the field ($\beta_e > 1$). Thus $L_H > L_A > L_O$ in the lower ionosphere. Physically this means that in the rest frame of the neutral, the current density is almost entirely due to electrons as is the Lorentz force $\J\times\B/c$. Thus there must be an electric field $\E \approx - \J\times\B/e\,n_e\,c$ in this frame that prevents the electron from accelerating, since collisions are ineffective ($\beta_e > 1$). As a result the {\it Ohm{}\'s Law} in the frame of the neutrals becomes $\sigma_H\,\E \approx \hB\times\J\,,$ where $\sigma_H \approx e\,n_e\,c/B$ is the Hall conductivity which is related to the Hall diffusivity,  $\eta_H = c^2/\left(4\,\pi\,\sigma_H\right)$. The parallel electric field component is much smaller than the perpendicular component because the Ohmic conductivity $\sigma_O = \beta_e\,\sigma_H$ is much larger than $\sigma_H$. Therefore, Hall is the dominant diffusion mechanism at the lower altitude. The parameter $\beta_e$ increases from $1$ in D layer ($60—-90\,\mbox{km}$) and reaches $10^4$ at $200\,\mbox{km}$. It is clear from the top panel of the Fig.~(\ref{fig:DHV}) that the Hall is the dominant diffusion in the lower and mid altitude of the ionosphere. Therefore, the magnetic flux is frozen in the E region and F region of the ionosphere in a frame moving with the combined velocity $\v+\vB$. Further the Hall scale, Eq.~(\ref{eq:LH}) which depends on the fractional ionization $X_e$
\bq
L_H \cong X_e^{-1/2}\,\delta_i\,,          
\label{eq:HSL}
\eq 
where $\delta_i = v_{A\,i} / \omega_{ci}$ is the ion--inertial scale, is very large for the typical values of $X_e \sim 10^{-9}—-10^{-4}$ in the lower and middle ionosphere, it is clear why the Hall diffusion is the dominant diffusion mechanism of the magnetic field. However, above mid—-altitude, when $\beta_i > 1$ the ambipolar diffusion will dominate the Hall resulting in the attenuation of the waves.  We also see from the figure that beyond the mid altitude Ohm diffusion is much smaller than the ambipolar, and as can be seen from Eq.~(\ref{eq:ets}) the Pedersen and Ambipolar diffusion become identical.  

Although the non—-ideal diffusion terms in the induction Eq.~(\ref{eq:ind1}) appear as a result of the frequent plasma—-neutral collision, their relative importance on the magnetic diffusion can only be gauged by comparing them to the advection term. To compare the advection term $\curl\left(\v\cross\B\right) \sim v\,B / L$ with the various diffusion terms in the induction Eq.~(\ref{eq:ind1}) we define magnetic Reynolds numbers $\mbox{Rm}\,\,,\mbox{Am}\,\,\&\,\, \mbox{Hm}$ as
\bq
\mbox{Rm} = \frac{v\,L}{\eta_O}\,\,,\,\mbox{Am} = \frac{v\,L}{\eta_A}\,\,\&\, \mbox{Hm} = \frac{v\,L}{\eta_H}\,.
\eq
Recall that the single fluid formulation of the partially ionised matter is valid for the frequencies smaller than the neutral—-ion collision frequency [Eq.~(\ref{eq:omega1})]. Thus, the bulk velocity is of the order of $ v_A^2 / L\,\nu_{ni}$. Clearly, the Reynolds numbers in a partially ionised fluid becomes  
\bq
\mbox{Rm} = \frac{v_A^2}{\nu_{ni}\,\eta_O}\,\,,\,\mbox{Am} = \frac{v_A^2}{\nu_{ni}\,\eta_A}\,\,\& \,\,\mbox{Hm} = \frac{v_A^2}{\nu_{ni}\,\eta_H}\,.
\eq
The variation of various Reynolds numbers against height is shown in the bottom panel of the Fig.~(\ref{fig:DHV}). It is clear from the figure that Ohm is lest important of the three diffusion mechanism. Thus $Rm>1$ after $110—-120\,\mbox{km}$. However, $Hm$ and $Am$ remains smaller than one throughout implying that both Hall and ambipolar diffusion will dominate the advection of the field. 
\begin{figure}
      \includegraphics[scale=0.45]{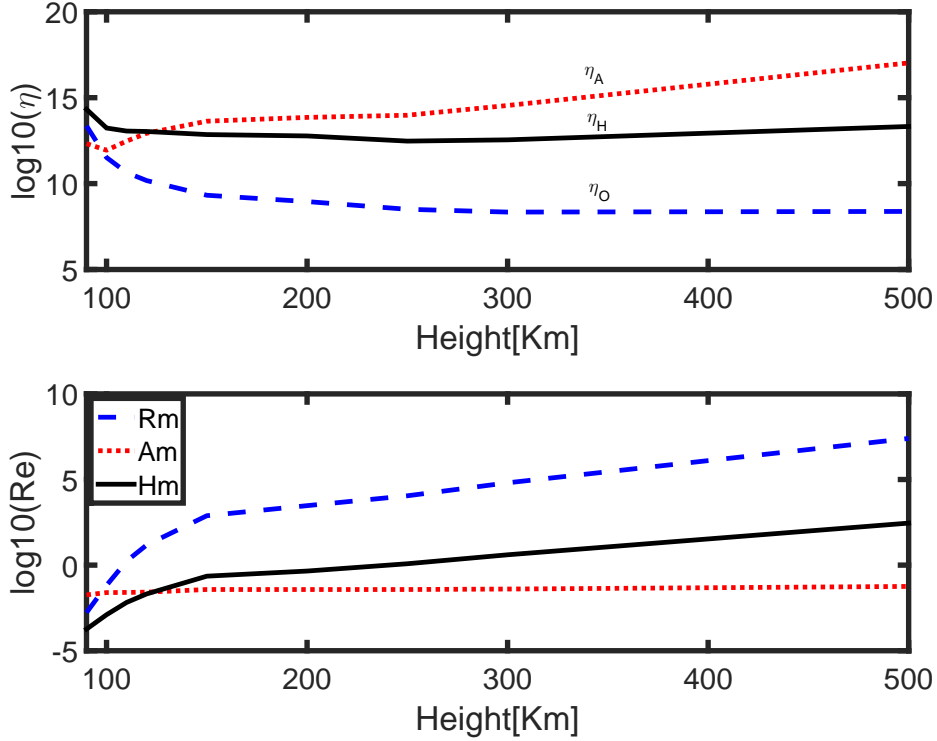}
\caption{ The variation of Hall (solid), ambipolar (dashed) and Ohm (dotted) diffusivities (top panel) and the variation of respective Reynolds numbers (bottom panel) in the ionosphere is shown in the above figure.}
\label{fig:DHV}
\end{figure}
 
Note that in the ionosphere literature [12\,,14], rather than dealing with the induction Eq.~(\ref{eq:ind1}) one often deals with the generalized Ohm{}\'s law 
\bq
\J= \sigma\,\Epa + \sigma_{P}\,\Epe + \sigma_H\,\hB\cross\Ep\,,
\label{eq:olw}
\eq
where $\Epa=\Ep\cdot\hB\,, \&\,\Epe \hB\cross\left(\Ep\cross\hB\right)$.  
The expression for various conductivities are well known [16].  Recall that the above equation is derived from the momentum equations (\ref{eeq})--(\ref{ieq}) by assuming that the fluid evolves on a slow (compared to the plasma—-neutral collision) time—-scale so that the drift of the charged particles through the neutral is determined by the instantaneous Lorentz force. Inverting Ohm{}\'s law, Eq.~(\ref{eq:olw}) gives
\bq
\Ep = \eta\,\Jpa + \eta_H\,\J\cross\hB + \eta_P\,\Jpe\,,
\label{eq:iolw}
\eq
where the perpendicular component of the current $\J$ refers to the orientation with respect to the ambient magnetic field
\bq
\Jpa = \left(\J\cdot\hB\right)\,\hB\,,\quad \Jpe = \J - \Jpa\,.
\eq
Taking curl of Eq.~(\ref{eq:iolw}) and making use of Maxwell{}\'s equation $c\,\curl\E = - \partial_t\,B$ we recover the induction equation (\ref{eq:ind1}).  The relationship between diffusivity and conductivity is well known [16]. We see that both these formulations are equivalent.

The energy equation (see the appendix for the derivation) in a partially ionized plasma becomes
\begin{eqnarray}
\frac{\partial}{\partial t} \varepsilon
  + \grad\cdot\left(\varepsilon\,\v\right)  
= - 
\grad\cdot\left[ P\,\v + \left(\v + 2\,\vB\right)\,\frac{B^2}{8\,\pi} 
- \left(\v + \vB\right)_{\parallel}\,\frac{B^2}{4\,\pi}
\right] 
\nonumber\\
- \eta_P\,\J_{\perp}^2\,,
\label{eq:egy}
\end{eqnarray}
where
\bq
\varepsilon = \frac{\rho\,v^2}{2} + \frac{P}{\gamma-1} + 
\frac{B^2}{8\,\pi}\,.
\eq
The energy equation for purely ambipolar case reduces to Eq.~(9) of Parker [17]. Above energy equation encompasses both the ambipolar and Ohm diffusions.

Note that in the absence of non—-ideal MHD terms, i.e. setting $\eta = \eta_P = \eta_H = 0$ or, $\vB = 0$, the set of Eqns.~(\ref{eq:continuity}), (\ref{eq:meq1}) and (\ref{eq:ind1}) and (\ref{eq:egy}) becomes identical to the ideal MHD equations, except now the plasma is partially ionized.  We know that the ion Larmour radius provides an implicit scale in the ideal MHD theory [18] and thus the validity of the ideal MHD requires that the characteristic length scale of a physical system under consideration should be larger than the Larmor radius $r_L = c_s / \omega_{ci}$. Here $c_s$ is the sound speed. In a partially ionized plasma this scale is$^{19}$ 
\bq
R_L = X_e^{-1}\,r_L\,. 
\label{eq:mlR}
\eq 
Clearly, the ideal MHD like description of a partially ionized plasma is valid when the characteristic scale length $L$ is larger than the modified Larmor radius $R_L$. Therefore, although the ideal MHD equations for both the fully and partially ionized plasmas looks similar, the scale of their validity is quite different. 
\section{Waves in the ionosphere}
In order to study the propagation of low frequency waves at some characteristic altitude in the ionosphere, we shall assume that the medium is locally homogeneous with no background flow. The local uniformity condition requires that the wavelength of fluctuations is much smaller than the inhomogeneity scales. To keep the analysis simple, rather than using energy Eq.~(\ref{eq:egy}), we shall close the set of equations by an isothermal equation of state $P = c_s^2\,\rho$. We note that bot the linear and nonlinear studies of the wave propagation has been carried out in Ref.~[9]. However, the focus of the present investigation is to highlight the role of various magnetic diffusion on the waves and thus is complementary to the previous investigation.  

After linearizing the Eqs.~(\ref{eq:continuity}),  (\ref{eq:meq1}) and (\ref{eq:ind1}) and Fourier analysing as 
$exp \left(\omega\,t + i \k\cdot\x\right)$ we get the following dispersion relation [PW08]
\begin{eqnarray}
\left\{\Big[\omega^2 - \left( v_A^2 + i\,\eta_A\,\omega\right)\,k^2 \,\cos^2\theta
- i\, \eta_O\,k^2\,\omega\Big] - \,k^2 \,\sin^2\theta \,
\left[\frac{\omega^2}{{\bar{\omega}}^2}\,v_A^2  + i\,\eta_A\,\omega \right] \right\} \times
\nonumber \\
\Big[\omega^2 - \left( v_A^2 + i\,\eta_A\,\omega\right)\,k^2 \,\cos^2\theta
- i\, \eta_O\,k^2\,\omega\Big]
 - \left(k^2\,\eta_H\right)^2\,\omega^2\,\cos^2\theta = 0\,.
\label{eq:DER}
\end{eqnarray}
Here $\theta = \left(\k\cdot\B\right)/k\,B$ is the angle between the wave vector $k$ and magnetic field $B$ and $\bar{\omega}^2 = \omega^2 - k^2\,c_s^2$.

When the wave is propagating along the magnetic field ($\theta = 0$), the dispersion relation reduces to 
\bq
\omega^2 - i\,k^2\left(\eta_A + \eta\right)\,\omega
- \omega_A^2 = \pm \eta_H\,k^2\,\omega\,.
\label{eq:WN}
\eq

The plus and minus sign describes the left and right circularly polarized waves. The Hall diffusion introduces the handedness in the system. In the low frequency ($\omega \ll \omega_A$) limit neglecting $\omega^2$ term in Eq.~(\ref{eq:WN}) gives
\begin{eqnarray}
Re[\omega] =  \frac{\omega_H }{ 1 + \left(\beta_e^{-1} +  D^2\,\beta_i\right)^2}\,,\nonumber\\    
Im[\omega] = \frac{\omega_H\,\left(\beta_e^{-1} + D^2\,\beta_i\right)}{ 1 + \left(\beta_e^{-1} +  D^2\,\beta_i\right)^2}\,.
\label{eq:ICM}
\end{eqnarray}
which describes the modified ion--cyclotron waves. Note that since at the lower and mid altitude where $D\approx 1$ the electrons are well coupled to the magnetic field ($\beta_e \gg 1$) whereas ions are not ($\beta_i \lesssim 1$) the waves will propagate almost undamped as $Im[\omega] \approx 0$. However since
\bq
\left(\frac{Re[\omega]}{\omega_A}\right) \approx \left(\frac{\omega_H}{\omega_A}\right) \equiv \left(\frac{1}{k\,L_H}\right)\,,
\eq 
we see that in the Hall regime ($k\,L_H >>1$) the ions hardly respond to the magnetic fluctuations. This explains why the magnetic flux is frozen in the electron fluid only in the Hall regime.

In the high frequency limit $\omega_A \ll \omega$ limit and assuming $\eta_O = 0$, we get
\bq
\frac{Re[\omega]}{\omega_A}=k\,L_H\,,\mbox{and}\,,    
\frac{Im[\omega]}{\omega_A} = \beta_i\,k\,L_H\,,
\label{eq:ICM1}
\eq
which describes the whistler waves in a partially ionized medium. Since ions are unmagnetized below mid altitude, the whistler propagates undamped at these altitudes. However, once $\beta_i \gtrsim$, the whistlers will suffer damping in the middle and upper ionosphere.

Since above the mid-—altitude ambipolar scale length is the largest Eq.~(\ref{eq:LH}) since $\beta_i > 1$, the ambipolar becomes the dominant diffusion in the plasma [Fig.~(\ref{fig:DHV})], and the dispersion relation Eq.~(\ref{eq:WN}) after neglecting both the Hall and Ohm terms becomes 
\[
\frac{Re[\omega]}{\omega_A} \approx \pm \left[ 1 - \frac{1}{2}\,\left( \frac{k\,\eta_A }{v_A}\right)^2 \right]^{1/2}\,,
\]
\begin{equation}
\frac{Im[\omega]}{\omega_A} \approx k\,L_A \equiv \beta_i\,k\,L_H\,.
\label{eq:art}
\end{equation}

\begin{figure}
      \includegraphics[scale=0.45]{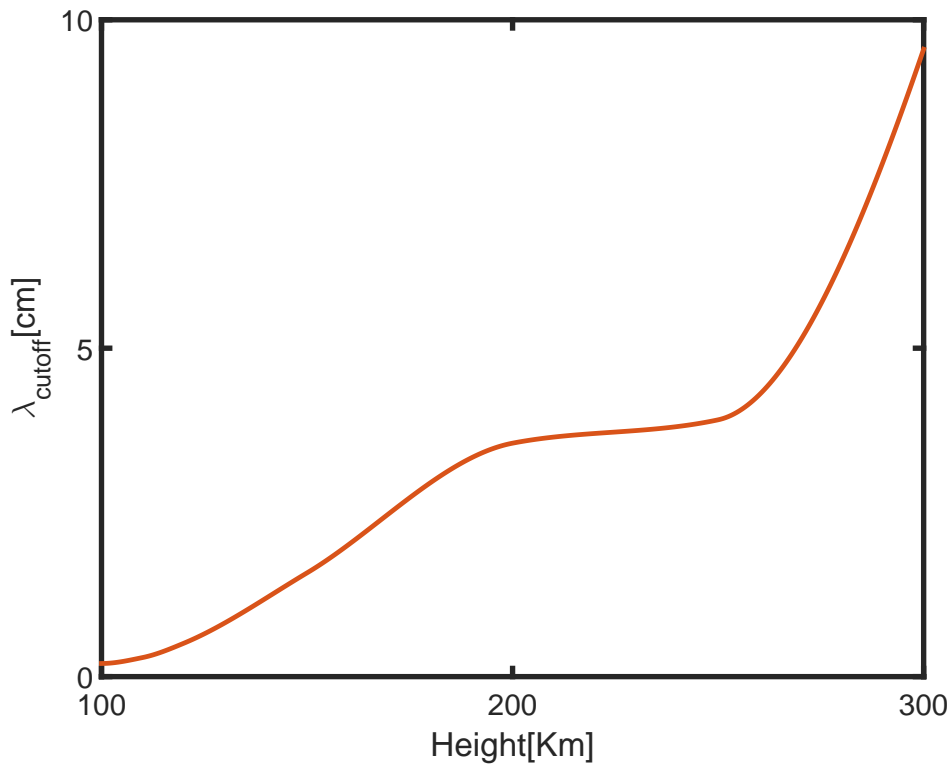}
\caption{The typical cut—off wavelength against height in the ionosphere.}
\label{fig:cof}
\end{figure}
It is well known that the waves are damped in the weakly ionized collisional medium [20]. The damping of the waves is not only dependent on the ion-neutral collision frequency but also on the ratio of the neutral to the bulk mass densities $D$. Clearly, only long wavelength signal can propagate undamped in the collision dominated ionosphere, i.e. when the wave frequencies are smaller than the 'effective' neutral-ion collision frequency, i.e. $k\,L_A \ll 1$. Only signals of certain wavelength will disappear due to the collisional dissipation, and the waves with wavelengths $\lambda$ exceeding
\bq
\lambda_{\mbox{cut}-\mbox{off}} \simeq \sqrt{2}\,\pi\, X_e^{-1}\,\left(\frac{v_A}{\nu_{in}}\right)\,,
\eq  
propagates in the medium. Since the magnetic restoring force due to the field deformation of wavelength $\lambda$ acts only on the plasma particles, the neutrals at a distance $\lambda$ apart can respond simultaneously to this restoring force only if the communication time between the neutrals ($\sim \lambda /v_A$) is smaller than the 'effective' collision time $t_{c} \sim X_e^{1/2}/\nu_{in}$. Only when $t_{c}$ exceeds the \alf crossing time $\sim \lambda /v_A$, the wave in the medium will damp. We see from the Fig.~(\ref{fig:cof}) that any wave with wavelength larger than few centimetre will propagate undamped in the lower and middle ionosphere. In the upper ionosphere $\lambda_{\mbox{cutoff}}$ is few meters. However, since ambipolar diffusion dominates Hall in the upper ionosphere, the waves will undergo significant damping. 

For waves propagating transverse to the magnetic field, i.e. assuming $\theta = \pi/2$, the dispersion relation Eq.~(\ref{eq:DER}) nicely factorizes with one of the factors giving $\omega = 0$ and a purely damped mode
\bq
\omega = i\,k^2\,\eta_O\,.
\eq
and the other factor gives magnetosonic 
\bq
\omega^2 = k^2\,\left(c_s^2 + v_A^2\right)\,,
\eq
and a purely damped mode
\bq
\omega = i\,k^2\,\eta_P\,.
\label{eq:drt}
\eq
When the direction of wave propagation is oblique to the magnetic field, the magnetosonic mode will split into fast and slow modes.

\section{Discussion and summary}
The multi--fluid description of highly collisional ionosphere can be reduced to single-fluid MHD—-like description in the low—-frequency limit. By \alf waves in the ionosphere we often imply waves in the fully ionized, ideal MHD fluid where the ion inertia balances the field deformation. Frequency of these waves is often too high and thus may undergo severe damping. However, partially ionized, collision-dominated ionosphere also supports low-frequency \alf wave which is caused by a balance between the bulk fluid inertia and the deformation of the magnetic field. In the vanishing plasma inertia limit in the E region, it means that the inertia of the wave is solely due to the neutral particles. The collisional momentum exchange plays a crucial role in transferring the magnetic stresses to the neutrals. Therefore, in the low-frequency (in comparison with the neutral-ion collision frequency) limit, the low frequency waves in the ionosphere will propagates with very little damping.

We have assumed locally uniform medium and thus the fluctuation wavelength must be smaller than the inhomogeneity scale length. This assumption is quite restrictive in the Earth$\textquoteright$s ionosphere. Further, the vertical stratification of the medium has been neglected. This does not seem to be as restrictive as it appears. For example, in the Hall diffusion dominated E region of the ionosphere, neglecting Ohm and ambipolar diffusion and assuming vertical stratification of the physical quantities in the presence of a uniform vertical field, the system admits an exact solution 
\begin{eqnarray}
B_x = B_0\,\cos\phi\,,\quad B_y = B_0\,\sin\phi\,,
\nonumber\\
V_x = V_0\,\cos\phi\,,\quad V_y = V_0\,\sin\phi\,,
\end{eqnarray}
which we recognise as the finite—-amplitude circularly polarized \alf waves with the phase $\phi = k_0\,z-\omega_0\,t$. Here the wavenumber $k_0$ and frequency $\omega_0$ are related by the following dispersion relation
\bq
\omega_0^2 = k^2\,v_A^2\,\left(1\pm\frac{\omega_0}{\omega_H}\right)\,,
\eq
which in the limiting cases describes the modified ion—-cyclotron [Eq.~(\ref{eq:ICM})] and whistler  waves [Eq.~(\ref{eq:ICM1})] in the short wavelength ($\omega_H \ll k_0\,v_A$) limit,  and \alf wave ($\omega_0 = k_0\,v_A$) in the long ($k_0\,v_A \ll \omega_H$) limit. Clearly, vertical stratification in the Hall dominated E-—region leads to a similar result as the local linear analysis. Using existing analytical tools, the weakly nonlinear, weakly dispersive processes can be easily investigated in the ionosphere [9]. 

Owing to the vast spatial and temporal scales involved in the ionosphere, the simplified local analysis is quite restricted. However, it provides important insight to the magnetosphere—-ionosphere coupling. The low frequency waves which is excited in the lower ionosphere due to the {\it loading} of neutral inertia on the field line have very large wavelength and thus couples to the magnetosphere. The frequency of these waves are below few Hz. For example, the fundamental frequencies at which magnetosphere resonates is $1.3$ and $1.9\,\mbox{mHz}$ [21]. Equating this with the whistler speed at $100\,\mbox{km}$ where  $v_A = 4\times10^3\,\mbox{cm}/\mbox{s}$ gives $\lambda \gtrsim 1000\,\mbox{km}$. This crude estimate suggests that the generation of low—-frequency waves in the E-—region of the ionosphere probably could be the cause of the magnetospheric resonant tuning. However, only a proper detailed analysis of the problem can lend credence to such a conjecture.       

Large wind shears have been observed in the mesosphere and lower thermosphere region over a wide range of altitudes, longitudes, seasons, and local times for the past five decades [22].  The wind shears often exceeds $70\,\mbox{m}/ \mbox{s} / \mbox{km}$  and can occasionally reach $100\,\mbox{m}/ \mbox{s} / \mbox{km}$.  The presence of winds and wind shears in the mesosphere and lower thermosphere region causes Kelvin--Helmholtz instability. However, the role of magnetic diffusion on the wind shear and thus on the onset of such an instability is unclear. Such an investigation may improve our understanding of irregular structure formation in the lower ionosphere.

To summarize, we have given a single fluid MHD-like formulation of the highly collisional partially ionized ionospheric plasma which with the increasing altitude reduces to the usual MHD description of the collisonless magnetosphere. Although similar formulation with zero plasma inertia and without energy equation was given in ref.~[9] for the ionosphere, they did not elaborate on the role of various diffusive scales in the ionosphere. Further, the limitations of such a formulation  
though implied was not discussed explicitly. The present framework provides a unified non-—ideal framework for the description of the ionosphere—-magnetosphere coupling. The non—-ideal MHD effects which depends are dominant in the ionosphere become negligible with increasing height. This results in a seamless transition from the single—-fluid MHD like description of the partially ionized ionosphere to the single-—fluid MHD description of the fully ionized magnetosphere.

{\bf{ACKNOWLEDGEMENTS}}
It is my great pleasure to acknowledge numerous discussions with Prof. Mark Wardle on the subject of partially ionized plasmas in space and astrophysics.

\newpage

\begin{center}
{\bf{Appendix}}
\end{center}

We shall assume the following thermodynamic relationship $P\,\rho^{-\gamma} = \mbox{const.}$, i.e.
\bq
\frac{\partial P}{\partial t} +  \left(\v\cdot \grad\right) P = - \gamma\,P \,\left(\grad\,\cdot\v\right)\,.
\label{eq:press}
\eq
  
In order to derive the energy equation, we multiply Eqn.~(\ref{eq:meq1}) scalerly with $\v$. Note that since 
\bq
\left(\v\cdot \grad\right)\v = \nabla \left( \frac{v^2}{2}\right) + \left(\curl\v\times\v\right)\,,
\eq
the left hand side of the momentum equation, combined with the continuity equation becomes
\bq
\frac{\partial }{\partial t} \left(\frac{\rho\,v^2}{2}\right) + \grad\cdot\left(\frac{\rho\,v^2}{2}\,\v\right)\,.
\eq

Next, lets evaluate $\v\cdot \grad P$. Note that
\begin{eqnarray}
\frac{\partial P}{\partial t} = - \left(\v\cdot \grad\right) P - \gamma\,P \,\left(\grad\,\cdot\v\right) 
\equiv \left(\gamma - 1 \right) \left(\v\cdot \grad\right) P - \gamma \grad \left( P\,\v\right)\,,
\end{eqnarray}
which gives 
\bq
\left(\v\cdot \grad\right) P = \frac{\partial }{\partial t} \left(\frac{P}{\gamma-– 1}\right) + \left(\frac{\gamma}{\gamma-1}\right)\,\grad\cdot\left(P\,\v\right)\,.
\label{eq:pt}
\eq
Lastly the Lorentz term dotted with $\v$ in the momentum equation is 
\begin{eqnarray}
\v\cdot\left(\curl\B\right)\cross\B = - \left(\v\cross\B\right)\cdot \left(\curl\B\right)\nonumber \\
 = - \left\{ \left[\left(\v + \vB\right) \cross\B \right] - \vB\cross\B \right\}\cdot \left(\curl\B\right)\,. 
\end{eqnarray}
Since,
\begin{eqnarray}
c\,\E_{\perp} + \left(\v + \vB\right)\cross\B = 0\,\nonumber \\
\Rightarrow\,\quad
- \left\{ \left[\left(\v + \vB\right) \cross\B \right] - \vB\cross\B \right\}\cdot \left(\curl\B\right) \nonumber\\
=  c\,\E_{\perp}\cdot \left(\curl\B\right) + \vB\cross\B\cdot \left(\curl\B\right)
\end{eqnarray}
Note that
\bq
- \E_{\perp}\cdot \left(\curl\B\right) = c^{-1} \frac{\partial }{\partial t} \frac{B^2}{2} + \grad\cdot \left(\E\cross\B\right)\,.
\eq
Thus defining the total energy 
\bq
\varepsilon = \frac{\rho\,v^2}{2} + \frac{P}{\gamma-1} + 
\frac{B^2}{8\,\pi}\,,
\eq
the energy euation can be written as
\begin{eqnarray}
\frac{\partial}{\partial t} \varepsilon = 
- \grad\cdot\left[
\left(\frac{\rho\,v^2}{2} + \frac{\gamma\,P}{\gamma-– 1} \right)\,
\v + \frac{c\,\E\cross\B}{4\,\pi} \right] 
\nonumber\\
 + \left(\vB\cross\B\right)\cdot \left(\curl\B\right)\,,
\end{eqnarray}
where
Making us of Eq.~(\ref{eq:md0}) we see that 
$$
\left(\vB\cross\B\right) = - \eta_P\,\left(\grad\cross\B\right)_{\perp} - \eta_H\,\left(\grad\cross\B\right)_{\perp}\cross\hB\,, 
$$
thus the energy equation becomes
\bq\frac{\partial}{\partial t} \varepsilon = 
- \grad\cdot\left[
\left(\frac{\rho\,v^2}{2} + \frac{\gamma\,P}{\gamma-– 1} \right)\,
\v + \frac{c\,\E\cross\B}{4\,\pi} \right] 
+ \eta_P\,\left(\grad\cross\B\right)_{\perp}^{2}\,.
\eq
The energy equation can be written in a slightly different form noting that the Poynting flux 
\bq
\frac{c\,\E\cross\B}{4\,\pi} = \frac{1}{4\,\pi}\left[\left(\v+\vB\right)\,B^2 - \left(\v+\vB \cdot \B\right)\,\B
\right]
\eq
and writing
\begin{eqnarray}
\frac{\gamma\,P\,\v}{\gamma-1} + \frac{c\,\E\cross\B}{4\,\pi} = \left(\frac{P}{\gamma-1} + 
\frac{B^2}{8\,\pi} 
\right)\,\v + P\,\v \nonumber\\ 
+ \left(\v + 2\,\vB\right)\,\frac{B^2}{8\,\pi} - \left(\v + \vB\right)_{\parallel}\,\frac{B^2}{4\,\pi}\,.
\end{eqnarray}
Thus the energy equation becomes

\bq
\frac{\partial}{\partial t} \varepsilon
  + \grad\cdot\left(\varepsilon\,\v\right)  
= - 
\grad\cdot\Big[ P\,\v + \left(\v + 2\,\vB\right)\,\frac{B^2}{8\,\pi} 
- \left(\v + \vB\right)_{\parallel}\,\frac{B^2}{4\,\pi}
\Big] - \eta_P\,\J_{\perp}^2
\eq

\newpage

\end{document}